\documentstyle[12pt,epsfig,amsmath]{article}
\newcommand {\be}{\begin{equation}}
\newcommand {\ee}{\end{equation}}
\newcommand {\ba}{\begin{eqnarray}}
\newcommand {\ea}{\end{eqnarray}}
\begin{document}

\def \a'{\alpha'}
\baselineskip 0.65 cm
\begin{flushright}
\ \today
\end{flushright}

\begin{center}{\large
{\bf The Impact of Kaluza-Klein Excited W Boson on the Single Top
at LHC and Comparison with other Models }} {\vskip 0.5 cm} {\bf
Seyed Yaser Ayazi and Mojtaba Mohammadi Najafabadi}{\vskip 0.5 cm
} School of Particles and Accelerators, Institute for Research in
Fundamental Sciences (IPM), P.O. Box 19395-5531, Tehran, Iran
\end{center}

\begin{abstract}
We study the s-channel single top quark production at the LHC in
the context of extra dimension theories, including the
Kaluza-Klein (KK) decomposition. It is shown that the presence of
the first KK excitation of $W$ gauge boson can reduce the total
cross section of s-channel single top production considerably  if
$M_{W_{KK}}\sim2.2~ \rm TeV$ ($3.5~ \rm TeV$) for $7~\rm TeV$
($14~\rm TeV$) in proton-proton collisions. Then the results will
be compared with the impacts of other beyond Standard Model (SM)
theories on the cross section of single top s-channel. The
possibility of distinguishing different models via their effects
on the production cross section of the s-channel is discussed.

\end{abstract}

\section{Introduction}
The top quark is the heaviest particle yet discovered and might
be the first place in which new physics effects could appear. The
properties of the top quark could reveal information regarding
flavor physics, the electroweak symmetry braking mechanism and
physics beyond the Standard Model(SM). For this reason, the
search for signal top production is one of the major goal of the
Tevatron and Large Hadron Collider (LHC)\cite{LHC}. The top quark
production at Tevatron has been extensively studied. So far no
hint for the beyond the SM has been detected in Tevatron\cite{Top
physics}. With the increase in the number of the top quark events
at the Tevatron, the experimental uncertainties are expected to
be further reduced. Thus, the comparison between the observed top
quark production properties and more precise theoretical
calculations will be a new probe for possible existence of the
new physics. At the LHC, top quarks are produced primarily via
two independent mechanisms: The dominant production mechanism is
the QCD pair production processes $q\overline{q}\rightarrow
t\overline{t}$, $gg\rightarrow t\overline{t}$ and second is single
top production. Single top quarks at the LHC are produced via SM
in three kinematically different channels. The s-channel $W^*$
production, $q\overline{q'}\rightarrow W^*\rightarrow t
\overline{b}$ \cite{schanel}, the t-channel W-exchange mode,
$bq\rightarrow t q'$ \cite{tchan} and the $tW$ production
\cite{tw}. Since the cross section of single top production is
smaller than the cross section of $t\overline{t}$ production and
the final state signals suffer from large background events, the
observation of the single top events is even more challenging
than $t\overline{t}$. It is expected that sufficient integrated
luminosity and improved method of analysis will eventually
achieve detection of single top events at the LHC.

The process $q\overline{q'}\rightarrow W^*\rightarrow t
\overline{b}$, compare to the single top production via t-channel
can be reliably predicted and theoretical uncertainty in the
cross section is only about a few percent \cite{Heinson}. The
statistical uncertainty in the measurement of cross section for
this process at the LHC will be about $5.4\%$ with an integrated
luminosity of $30 fb^{-1}$\cite{1999fr}. At the leading order,
the cross sections of production  for all three processes are
proportional to the Cabibo-Kabayashi-Maskawa (CKM) matrix
elements. However, the LHC can precisely measure single top
production cross sections, and the CKM matrix elements could be
measured down to the less than one percent error at the ATLAS
detector \cite{1999fr}. Therefore, it is worthwhile to
investigate the effects of physics beyond SM on single top quark
production. Search for new physics via single top production has
been already investigated in \cite{BSM,Boos:2006xe,Tait:2000sh}.

In this paper, we study the effects of extra dimension theories
on the single top production via $q\overline{q'}\rightarrow
W^*\rightarrow t\overline{b}$ at the LHC. In extra dimension
theories, if the gauge fields of the SM propagate in the
 bulk of the extra dimension then they will have Kaluza-Klein (KK) excitations
that can couple to the SM fermions. The masses of KK excited
gauge boson will be proportional to inverse radius of
compactification. If the inverse of radius of compactification is
in order of $\rm TeV$ scale or less, phenomenological effects of
these states in ongoing colliders will be remarkable \cite{BSM}.
There are various experimental bounds on these masses which arise
from precision electroweak tests, cosmological constrains and low
energy observables \cite{constraints}. The lower bound on the
masses of the first excited of gauge boson obtained from $\rm W'$
searches in $p\overline{p}$ collision at Tevatron indicate that
they must lie above $\simeq915~\rm GeV$ \cite{Pangilinan:2010zz}.
If the fundamental Plank scale as a cutoff of the SM physics is at
around few $\rm TeV$, We expect that there will be
higher-dimensional operators suppressed by the cutoff scale. For
operators which respect symmetries of SM, the constraints come
from the electroweak precision tests. For example, Higgs mass is
limited by constraints which come from the requirements of vacuum
stability. It is shown \cite{Datta:1999dw} that the effects of
the higher dimensional operators on the Higgs mass are
significant. However, in theories with extra dimensions,
phenomenological constraints from CP violations, $\rm FCNC$ and
electroweak precision measurements give a bound on cutoff in
order of $\sim~10~\rm TeV$ \cite{Hall:1999fe, Banks:1999eg}. In
all of these analyses, it is assumed that the only new physics
beyond SM come from the KK excitations of SM fields. Note that
the gravity induced processes will affect on electroweak
observables and change lower bound on masses of KK excitations of
SM fields but will not significantly affect on single top
production \cite{gravity}. For this reason, we ignore the gravity
effects in our analysis. As it is mentioned above, precision
electroweak fits can significantly affect on lower bound of scale
of KK excitations fields ($M_c$). In the literature, it has been
shown \cite{pt} that these bounds are at the order of $\rm TeV$
scale. In light of these studies, we assumed that lower bound on
KK excitations of gauge fields is at the order of $\rm TeV$ scale.

In this paper, we study the contribution of the first excitation
KK mode of $W$ which denoted by $W_{KK}$, on the s-channel single
top production at LHC. It should be noted that for the high mass
region of the $W'$, its contribution  to the t-channel
 and $tW$-channel cross section are very small to be observed \cite{Boos:2006xe}.
The $W'$ in t-channel process is space-like
 and in tW-channel is real but the involved $W$ boson in the
 s-channel is time-like.

The rest of this paper is organized as follows: In the next
section, we summarize the effect of excited KK $W$ state on the
single top production. In section~3 we first evaluate the effect
of excited KK mode of $W$ on the s-channel single top production
at LHC and compare it with  results which arise from other
alternative new physics models. In this section, we discuss the
degeneracies between these models and survey the possibility of
solving this degeneracies. Our conclusions are given in section~4.

\section{KK excited W and its effects at LHC}
In this section, we briefly describe the model and study effects
of KK excited $W$ on single top production at LHC.

In this paper, we consider a simple extension of the SM to 5
dimension(5D) which was assumed as a part of more fundamental
underlying theory \cite{basic}. Note that gauge theories in more
than 4-dimensions are non-renormalizable and should be treated as
low energy effective theories below some cutoff
$\Lambda$\cite{Cheng:2010pt}. As SM fermions and Higgs carry gauge
quantum numbers, they cannot propagate in extra dimensions unless
the corresponding gauge fields also propagate in extra
dimensions. On the other hand, SM fermions and Higgs may still be
localized in 4 dimensions even if gauge fields propagate in extra
dimensions. In this paper, we suppose SM gauge fields propagate
in extra dimensions while fermions and Higgs live on 3-brane.
Another case which all SM fields live in the same extra
dimensions are strongly constrained by electroweak precision data
\cite{Cheng:2010pt, Appelquist:2000nn}. As is mentioned earlier,
we do not consider gravity induced processes in our analysis.
Because they have negligible contributions.

In this model, we consider one extra dimension compactified on a
circle with radius R. The coordinate are denoted as
$x_M=(x_{\mu},y)$, where $M=0,1,2,3,5$, $\mu=0,1,2,3$ and $y=5$
is the coordinate in the direction of the extra dimension. The
compactification means that the points $y$ and $y+2\pi R$ are
identified. As it is mentioned, we suppose that  SM gauge fields
live in 5D and so we can expand gauge fields with a Fourier
decomposition along with the compact dimension,
\begin{eqnarray}
\Phi_+(x_\mu,y)&=&\sum^{\infty}_{n=0}
\cos\frac{ny}{R}\Phi^{(n)}_+(x_\mu)\, ,\nonumber\\
\Phi_-(x_\mu,y)&=&\sum^{\infty}_{n=1}
\sin\frac{ny}{R}\Phi^{(n)}_-(x_\mu)\, , \label{fourier}
\end{eqnarray}
note that the fifth dimension $y$ is compactified on the orbifold
$S^1/Z_2$ with two 4D boundaries at $y=0$ and at $y=\pi R$. In
the above formula $\Phi^{n}_{\pm}$ are the KK excitations of the
5D gauge field and the gauge fields have defined to be even or
odd under the $Z_2$-parity,~$\rm i.e$
$\Phi_{\pm}(y)=\pm\Phi_{\pm}(-y)$. After integrating over the
fifth dimension, zero modes contain a 4D gauge field. In this
model, SM gauge fields live in 5D bulk, while the SM fermions and
Higgs doublets, can either live in the bulk or be localized on
the 4D boundaries. As a result, the fermions and Higgs doublet
couple to the KK excited gauge bosons only if they are localized
on the 4D boundaries. After integrating over the fifth dimension,
the effective four dimensional Lagrangian for charged electroweak
sector can be obtained,
\begin{equation}
{\cal{L}}^{ch}=\sum_{a=1}^2 {\cal{L}}_a^{ch}
\end{equation}
where
\begin{eqnarray}
{\cal{L}}_{a}^{ch}&=& \frac{1}{2}m^{2}_W W_a\cdot W_a
+\frac{1}{2}M_c^2\sum_{n=1}^\infty \, n^2 \, W_a^{(n)}\cdot
W_a^{(n)}
\nonumber\\
&-&g\,W_a\cdot J_a-g\,\sqrt{2} J^{KK}_a\cdot \sum_{n=1}^\infty
W_a^{(n)}\, ,
\end{eqnarray}
and $m^{2}_W=g^2v^2/2$, the weak angle $\theta$ is defined by
$e=g\, s_\theta=g'\, c_\theta$, while the currents are
\begin{eqnarray}
\label{currents} J_{a\mu}&=&\sum_\psi \bar{\psi}_L \gamma_\mu
\frac{\sigma_a}{2}\psi_L\, ,
\nonumber\\
J_{a\mu}^{KK}&=& \sum_\psi \varepsilon^{\psi_L}\bar{\psi}_L
\gamma_\mu \frac{\sigma_a}{2}\psi_L\, .
\end{eqnarray}
Note that for $\psi_L$ living in the bulk $\varepsilon^{\psi_L}$
is 0 and for fermions which are living in 4D will be 1.

The coupling of $KK$ excited $W$ to our model is determined in
terms of Fermi coupling $G_F$ up to corrections of
${\cal{O}}(m_Z^2/M_c^2)$ \cite{gravity}. For $M_c\sim1~\rm TeV$
the ${\cal{O}}(m_Z^2/M_c^2)$ effects are negligible for single
top production and therefore we do not consider these effects in
our calculations. We have ignored the mixing of the $W$ with
$W_{KK}$ which is also an ${\cal{O}}(m_Z^2/M_c^2)$ effects.
Therefore, in our model $W_{KK}$ decays only to SM particles. As
a result, to calculate the cross section of $pp\rightarrow
t\overline{b}X$, we only need to take into consideration the
couplings of the SM fermions to the KK excitations of the
electroweak gauge bosons and their unknown mass of $W_{KK}$.

The cross section of $pp\rightarrow t\overline{b}X$ is given by
\begin{eqnarray}
\sigma(pp\rightarrow t{\overline b} X) & = &\sum_{qq'} \int
dx_1dx_2[q(x_1){\overline q'}(x_2)+ q(x_2){\overline q'}(x_1)]
\widehat{\sigma}(q{\overline q'}\rightarrow t{\overline b}) . \
\end{eqnarray}
where $q(x_i)$ is structure function of $u$ or $c$ quarks and
$\overline{q'}(x_i)$ is structure function of $\overline{d}$ or
$\overline{s}$ quarks. $x_1$ and $x_2$ are the parton momentum
fractions. The partonic cross section takes the form
\cite{Boos:2006xe}:

\begin{eqnarray}
\label{formula_sigma} \widehat{\sigma}=\sum_{q,q'} \frac{\pi
\alpha_W^2}{6}|V_{tb}|^2 |V_{qq'}|^2 \frac{(\hat{s}-M_t^2)^2 (2
\hat{s}+M_t^2)}{\hat{s}^2} \left[
\frac{1}{(\hat{s}-m_W^2)^2+\gamma_W^2 m_W^2} + \right.
\\ \nonumber +
\frac{2(\hat{s}-m_W^2)(\hat{s}-M_{W_{KK}}^2)
+\gamma_W^2\Gamma_{W_{KK}}^2} {((\hat{s}-m_W^2)^2+\gamma_W^2
m_W^2)((\hat{s}-M_{W_{KK}}^2)^2 +\Gamma_{W_{KK}}^2 M_{W_{KK}}^2)}
+  \\ \nonumber \left. + \frac{1}
{(\hat{s}-M_{W_{KK}}^2)^2+\Gamma_{W_{KK}}^2 M_{W_{KK}}^2} \right]
\end{eqnarray}
where $\gamma_W$ is width of $W$ gauge boson, $\alpha_W =
g^2/(4\pi)$ and $\hat{s}=x_1 x_2S$ is parton center of mass
energy while $S$ is the $pp$ center of mass energy. Width of the
$W_{KK}$ is given by \cite{Tevatron}
\begin{eqnarray}
\Gamma_{W_{KK}} & \approx & \frac{2M_{W_{KK}}}{M_{W}}\gamma_W+
\frac{2M_{W_{KK}}}{3M_{W}}\gamma_W\cdot X ,\nonumber\\
X & = & ( 1-\frac{M_t^2}{M_{W_{KK}}^2}) (
1-\frac{M_t^2}{2M_{W_{KK}}^2} - \frac{M_t^4}{M_{W_{KK}}^4}). \
\end{eqnarray}

Note that $W_{KK}$ will have the same decays as the $W$ boson but
in addition it can also decays to top-bottom pair which is
kinematically forbidden for $W$ boson.

\section{Numerical results}
In this section, we study the effects of KK excitation of $W$
gauge boson on the cross section of production $pp\rightarrow
t\overline{b}X$ at LHC. We first discuss the experimental bounds
on mass of KK excited $W$. We then present our result and discuss
the possibility of detection of large extra dimensions at LHC via
single top. We compare our result with other new physics beyond
SM and their effects on production $pp\rightarrow t\overline{b}X$
at LHC. We then discuss the possibility of solving degeneracy
between different new physics models.

As it is discussed in previous section, there exists two kinds of
experimental constraints on the mass of KK excited $W$. First
kind of constraints arises from indirect search for large extra
dimensions. Model dependent limits can be placed on the masses of
KK excitation SM fields from precision electroweak test,
astrophysics ($e.g.$, star cooling), cosmology ($e.g$., expansion
rate of the universe) and low energy experiments ($e.g.$, CP
violation and FCNC processes)\cite{constraints,gravity,pt}.
Another kind of constraints comes from direct search of $W'$ at
existing colliders. A null result for a search of $W'$  in single
top production at Tevatron  indicate that mass of $W'$ must lie
above $915~\rm GeV$ \cite{Pangilinan:2010zz}. In our analysis, we
consider these bounds on mass of the first KK excitation mode of
$W$.
\begin{figure}
\begin{center}
\centerline{\vspace{-1.2cm}}
\centerline{\includegraphics[scale=0.4]{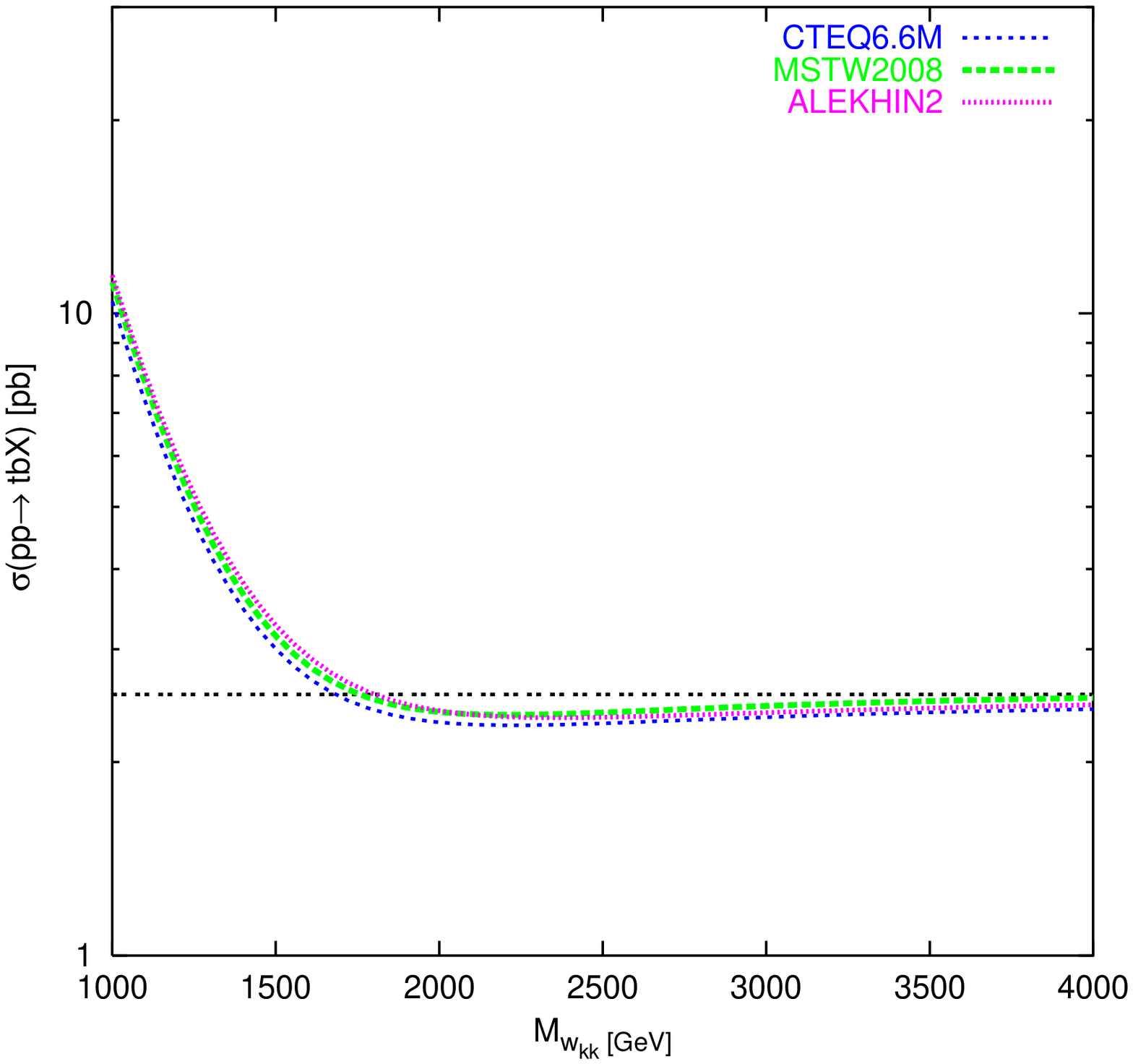}\hspace{5mm}\includegraphics[scale=0.4]{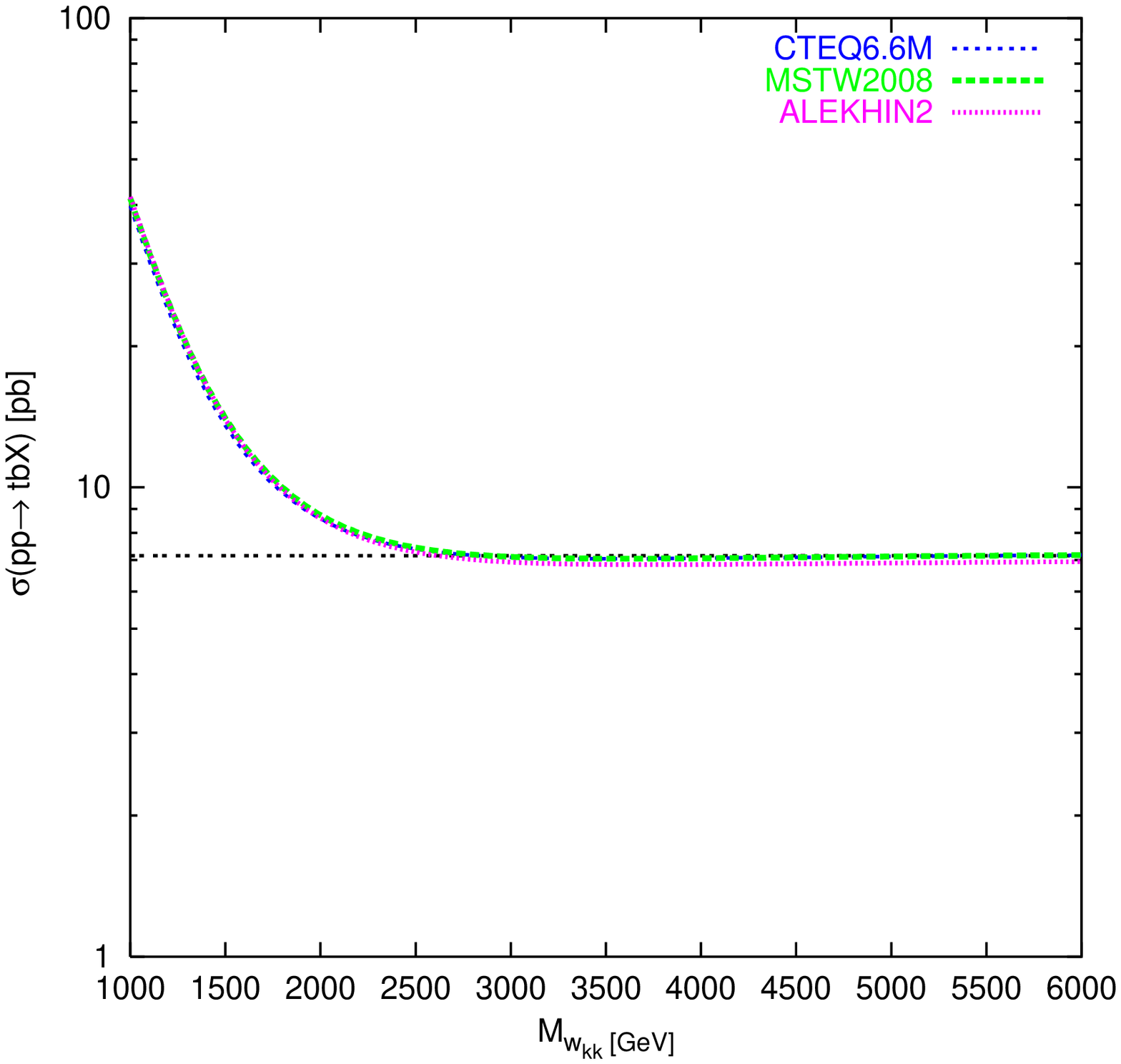}}
\centerline{\vspace{1.cm}\hspace{1cm}(a)\hspace{7cm}(b)}
\centerline{\vspace{-1.5cm}}
\end{center}
\caption{ a) $\sigma(pp\rightarrow t\overline{b}X)$ versus
$M_{W_{KK}}$, the mass of the first KK excitation mode of $W$. In
this figure, we have set $\sqrt{S}=7~ \rm TeV$. The horizontal
line at $2.55~\rm pb$ depicts SM prediction for this process at
$\sqrt{S}=7~\rm TeV$. The dotted (blue) curve, the dashed (green)
curve and small dotted (pink) curve respectively correspond to
CTEQ6.6M \cite{CTEQ}, MSTW2008 \cite{MSTW} and ALEKHIN2
\cite{Alekhin:2009vn} structure functions. b) Similar to Fig.~a
except that $\sqrt{S}=14~ \rm TeV$. The horizontal line at $7~\rm
pb$ depicts SM prediction for this process at $\sqrt{S}=14~\rm
TeV$.}\label{cross}
\end{figure}

In this paper, we study the effects of first KK excitation of $
W$ on s-channel single top production at LHC. In
\cite{Aad:2009wy}, it has been shown that in $7~\rm TeV$
collisions with small amount data, a $W'$ boson with a mass above
the present experimental limits could be found. For example, with
the integrated luminosity of $1~\rm fb^{-1}$, a $W'$ boson with
the mass of $3~\rm TeV$ can be reached.

\begin{figure}
\begin{center}
\centerline{\vspace{-1.2cm}}
\centerline{\includegraphics[scale=0.4]{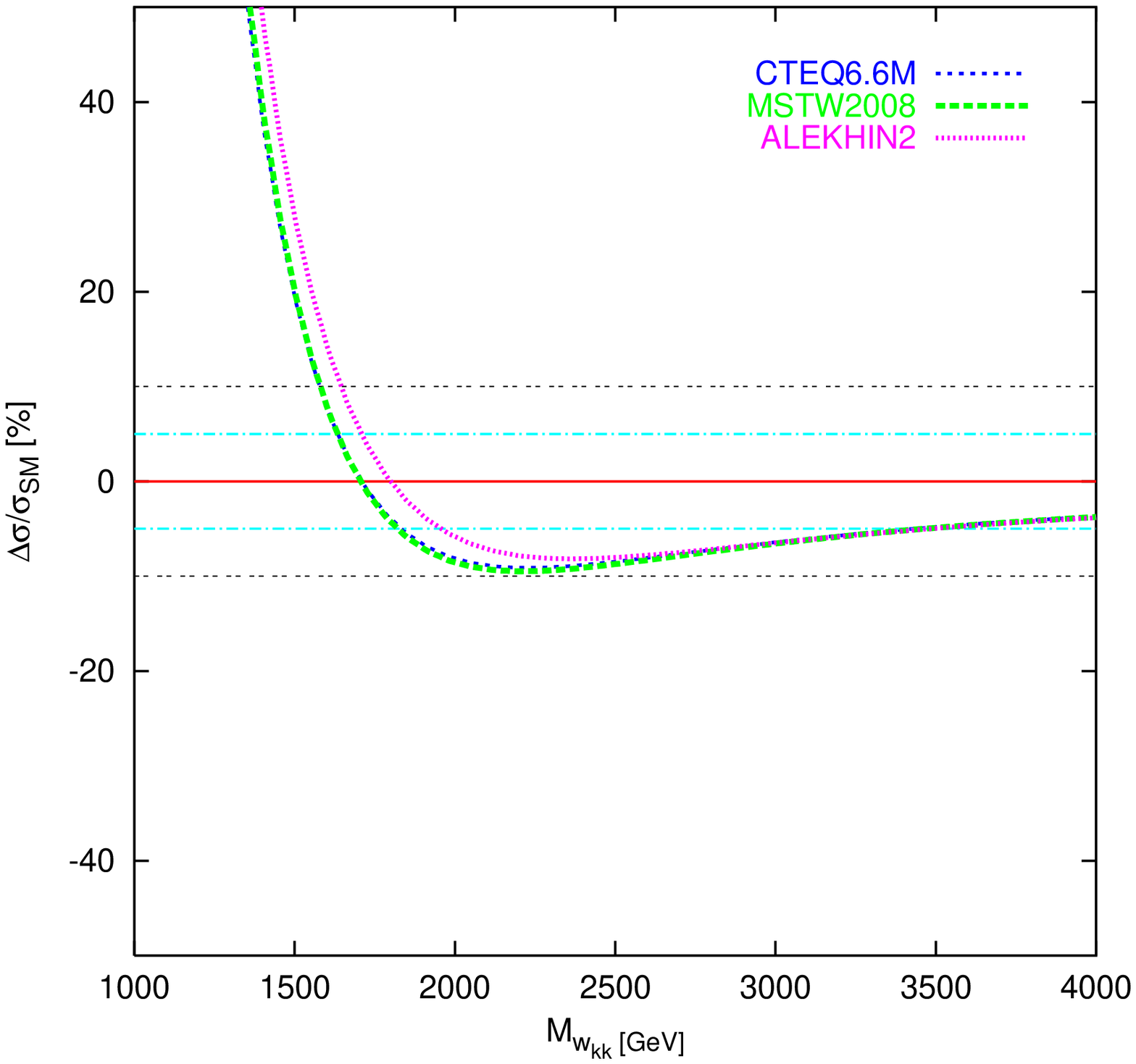}\hspace{5mm}\includegraphics[scale=0.4]{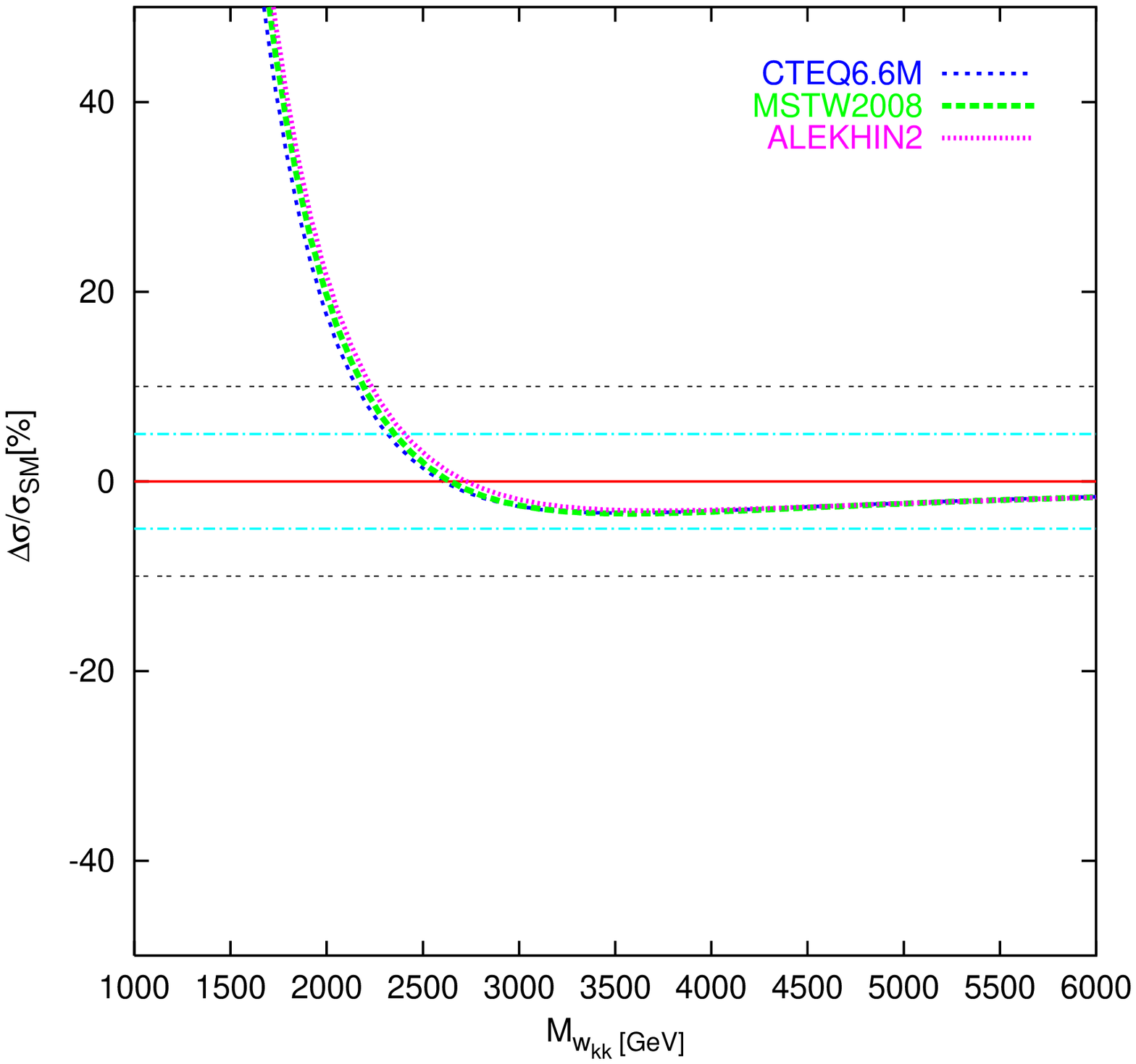}}
\centerline{\vspace{1cm}\hspace{1cm}(a)\hspace{7cm}(b)}
\centerline{\vspace{-1.5cm}}
\end{center}
\caption{ a) $\Delta\sigma(pp\rightarrow
t\overline{b}X)$/$\sigma_{SM}$ versus $M_{W_{KK}}$, the mass of
the first KK excitation mode of $W$. In this figure, we have set
$\sqrt{S}=7~ \rm TeV$. The dotted (blue) curve, the dashed
(green) curve and small dotted (pink) curve respectively
correspond to CTEQ6.6M \cite{CTEQ}, MSTW2008 \cite{MSTW} and
ALEKHIN2 \cite{Alekhin:2009vn} structure functions. b) Similar to
Fig.~a except that $\sqrt{S}=14~ \rm TeV$. The horizontal cyan
(gray) lines correspond to $\pm 5\%$($\pm 10\%$) uncertainty in
the measurement of $\sigma(pp\rightarrow
t\overline{b}X)$.}\label{deltacross}
\end{figure}

In Figure.~\ref{cross}, we display the cross section of
$pp\rightarrow t\overline{b}X$ versus $M_{W_{KK}}$ (the mass of
the first KK excitation mode of W). To calculate
$\sigma(pp\rightarrow t\overline{b}X)$, we have used the CTEQ6.6M
\cite{CTEQ}, MSTW2008 \cite{MSTW} and ALEKHIN2
\cite{Alekhin:2009vn} structure functions. The dotted (blue)
curve, the dashed (green) curve and small dotted (pink) curve
respectively correspond to CTEQ6.6M, MSTW2008 and ALEKHIN2
structure functions. As it is seen in this figure, different
structure functions do not affect on the value of
$\sigma(pp\rightarrow t\overline{b}X)$ more than $1\%$. In
Fig.(\ref{cross}.~a), we have set $\sqrt{S}=7~ \rm TeV$ and a
 SM cross section of $2.55~\rm pb$ for the
$\sigma(pp\rightarrow t\overline{b}X)$ is obtained. The
horizontal line in this figure depicts SM prediction for this
process. In Fig.(\ref{cross}.~b), as explained in the caption, we
have set $\sqrt{S}=14~ \rm TeV$. In the centre of mass of energy
$14~ \rm TeV$, leading order SM cross section of $pp\rightarrow
t\overline{b}X$ have been obtained $7~\rm pb$. The vertical line
at $915~ \rm GeV$ shows a lower bound on mass of KK excitation of
$ W$ which comes from null result for $W'$ search at Run IIa of
Tevatron \cite{Pangilinan:2010zz}. In Fig.\ref{cross}, there
exists a peak at areas that $M_{W_{KK}}$ is smaller than direct
lower bound. This is because of the momentum of the s-channel
resonance is time-like which leads to large interference with SM
amplitude. In Fig.~\ref{cross}, we observe that the presence  of
$W_{KK}$  can decrease total cross section. For better study of
this effect, we consider the relative change in cross section
which is given by
\begin{eqnarray}
 R=\frac{\Delta\sigma}{\sigma_{SM}}=\frac{\sigma-\sigma_{SM}}{\sigma_{SM}}. \
\end{eqnarray}
where $\sigma$ is total cross section in the presence of $W_{KK}$.
 The relative change in cross section of the single top production at
LHC are shown in Fig.~\ref{deltacross}. In this figure, we
display relative change in  cross section versus $M_{W_{KK}}$ for
$\sqrt{S}=7~\rm TeV$ and $\sqrt{S}=14~ \rm TeV$. The horizontal
cyan (gray) lines correspond to $\pm 5\%$($\pm 10\%$) uncertainty
in the  measurement of $\sigma(pp\rightarrow t\overline{b}X)$.
Fig.~\ref{deltacross} demonstrates that the effect of presence of
$W_{KK}$ for $M_{W_{KK}}>1500~\rm GeV$ can be destructive. Notice
that with increasing center of mass energy, deviation from SM
cross section decreases. As it was mentioned, the ATLAS detector,
can measure cross section for this process with statistical
uncertainty of $5.4\%$\cite{1999fr}. After integrated luminosity
of $30 fb^{-1}$, deviation  more than $5\%$ from SM cross section
can allude to beyond SM physics. Because the cross section of the
s-channel single top quark is proportional to the $|V_{tb}|^2$,
it provides the possibility to measure $V_{tb}$ element of CKM
matrix directly \cite{Beneke:2000hk}. Therefore, from the
measurement of the cross section of s-channel smaller than the SM
prediction one may conclude that $|V_{tb}|<1$ and this could be
interpreted as evidence for the presence of the new quark
generation mixed with the third family. However, the result of
this paper clearly shows that the measurement of the s-channel
cross section smaller than the SM prediction would not
necessarily indicate the evidence for extra family and it can be
due to a new charged heavy gauge boson. Now suppose that a
deviation of more than $5\%$ is observed in the production cross
section of s-channel single top. Can we distinguish between
different beyond SM theories by applying these results? In the
following, we study this problem with comparison of our result
with other beyond SM effects.

\begin{figure}
\begin{center}
\centerline{\vspace{-1.2cm}}
\centerline{\includegraphics[scale=0.65]{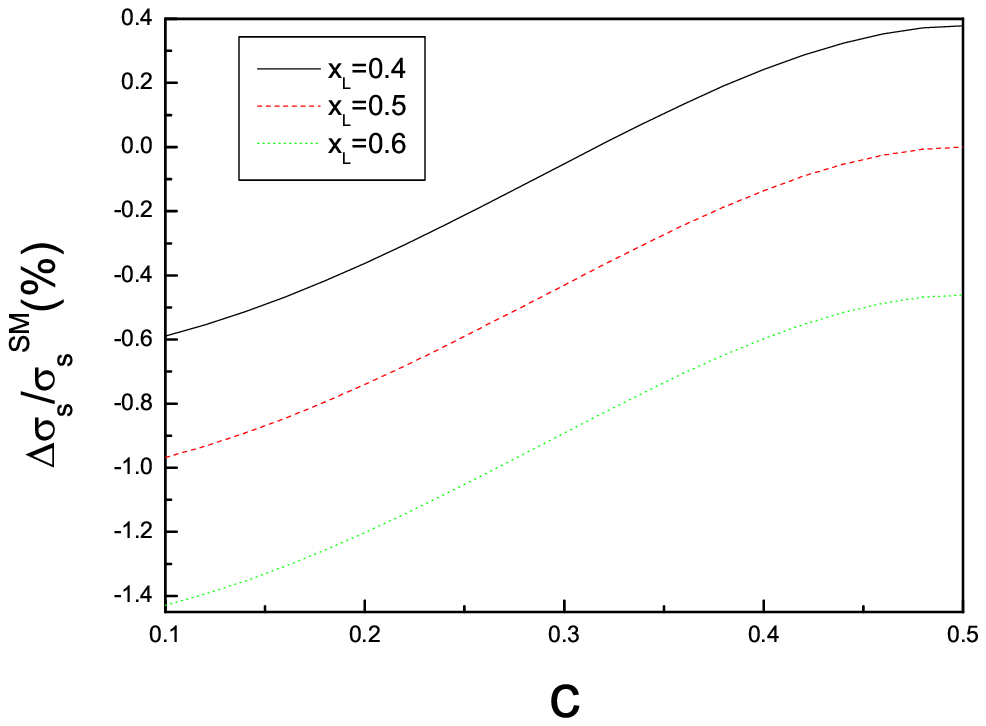}\hspace{1mm}\includegraphics[scale=0.67]{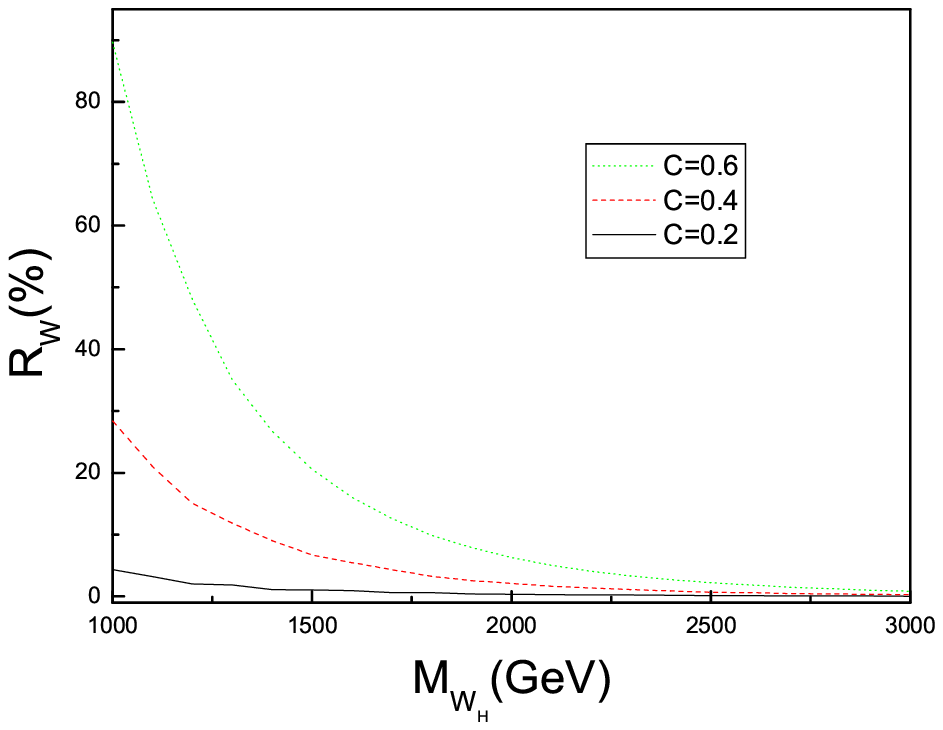}}
\centerline{\vspace{1.cm}\hspace{1cm}(a)\hspace{7cm}(b)}
\centerline{\vspace{-1.5cm}}
\end{center}
\caption{ a) $\Delta\sigma(pp\rightarrow
t\overline{b}X)$/$\sigma_{SM}$ versus mixing parameter $c$ in the
context of littlest Higgs models for $f=1~\rm TeV$ and different
values of the mixing parameter $x_L$. b)
$R_W=\Delta\sigma(pp\rightarrow t\overline{b}X)$/$\sigma_{SM}$
versus $M_{W_H}$ in the context of littlest Higgs models for three
values of the mixing parameters c. These curves have been
borrowed from Figs.~1 and 4 of \cite{Little
Higgs}.}\label{littlehigs}
\end{figure}

First we compare our result with Littlest Higgs Model(LH)
correction to s-channel single top cross section. The relative
corrections of the LH model to the cross section
$\sigma(pp\rightarrow t\overline{b}X)$ at the LHC are shown in
Fig.~\ref{littlehigs}. These curves have been borrowed from
Figs.~1 and 4 of \cite{Little Higgs}. In these figures, the
authors have been considered that $\Delta\sigma(pp\rightarrow
t\overline{b}X)=\sigma_{LH}-\sigma_{SM}$, $f=1.0~\rm TeV$ and
$c(s=\sqrt{1-c^{2}})$ is the mixing parameter between $SU(2)_{1}$
and $SU(2)_{2}$ gauge bosons and the mixing parameter
$x_{L}=\lambda_{1}^{2}/(\lambda_{1}^{2}+\lambda_{2}^{2})$ comes
from the mixing between the SM top quark $t$ and the vector-like
top quark $T$, in which $\lambda_{1}$ and $\lambda_{2}$ are the
Yukawa coupling parameters. The curves have been shown with three
values of the mixing parameters $x_{L}$ and $x_{\lambda}$.
Considering the constraints of the electroweak precision data on
these free parameters, they will be assumed $f\geq1 \rm TeV$,
$0.4\leq x_{L}\leq 0.6$ and $0<c\leq 0.5$ \cite{Little Higgs}.

From Fig.~(\ref{littlehigs}-a), we can see that the contributions
of the LH model to single top production are very smaller than
KK-excited $W$. The effect of LH model to s-channel single top
production might be destructive but at most up to $1.4\%$ for
$x_L=0.6$. This value is very smaller than precision of ATLAS
detector with $30 fb^{-1}$ data. Fig.~(\ref{littlehigs}-b) shows
the relative correction $R_W=\Delta\sigma(pp\rightarrow
t\overline{b}X)$/$\sigma_{SM}$ as a function of the mass new gauge
boson $W_H$ in the context littlest Higgs models for three values
of the mixing parameters $c$. The value of the relative correction
parameter $R_{W}$ is in the range of $ 1.5\%\leq R_{W}\leq 90\%
$. As it can be seen in this figure the effects of the new gauge
boson $W_{H}$ to the s-channel process for single top production
is always constructive  and we can distinguish between LH model
with our model for some areas in parameters space
($M_{W'}>2200~\rm GeV$).

\begin{figure}
\begin{center}
\centerline{\vspace{-1.2cm}}
\centerline{\includegraphics[scale=0.62]{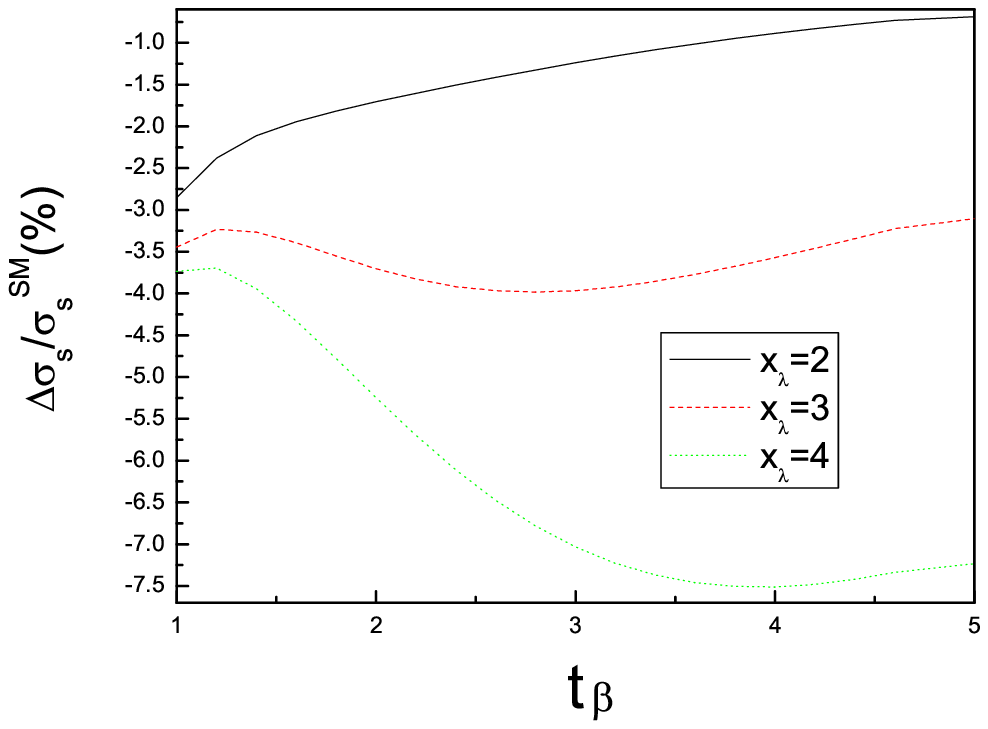}\hspace{1mm}\includegraphics[scale=0.7]{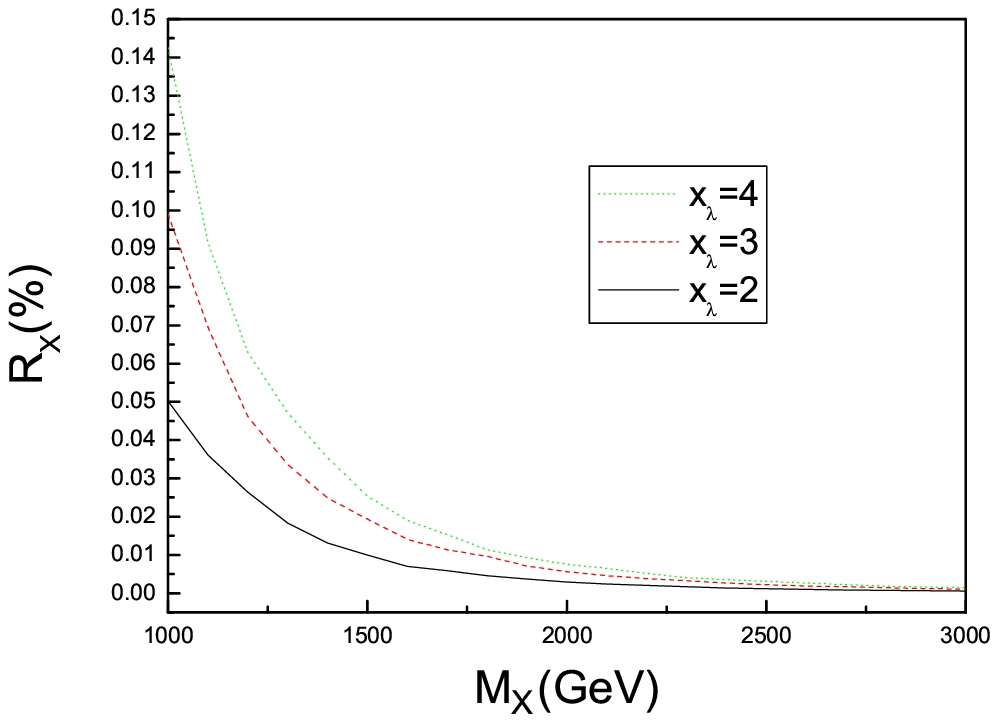}}
\centerline{\vspace{1.cm}\hspace{1cm}(a)\hspace{7cm}(b)}
\centerline{\vspace{-1.5cm}}
\end{center}
\caption{  $\Delta\sigma(pp\rightarrow
t\overline{b}X)$/$\sigma_{SM}$ versus the mixing parameter
$t_{\beta}$ in the context of $SU(3)$ simple group model for
$f=1~\rm TeV$ and different values of the mixing parameter
$x_{\lambda}$. b) $R_W=\Delta\sigma(pp\rightarrow
t\overline{b}X)$/$\sigma_{SM}$ versus $M_{X^-}$ in the context of
$SU(3)$ simple group model for $t_{\beta}=3$ and three values of
the mixing parameter $x_{\lambda}$. These curves have been
borrowed from Figs.~2 and 5 of \cite{Little Higgs}. }\label{su3}
\end{figure}

Another beyond SM theory which we are going to consider is $SU(3)$
simple group model. The relative corrections of the $SU(3)$ simple
group model to the cross section $\sigma(pp\rightarrow
t\overline{b}X)$ at the LHC has been shown in Fig.~\ref{su3}.
These curves have been borrowed from Figs.~2 and 5 of
\cite{Little Higgs}. In these figures, it is assumed that
$\Delta\sigma(pp\rightarrow t\overline{b}X)=\sigma_{\rm
SU(3)}-\sigma_{SM}$, $f=1.0~\rm TeV$ Where
$f=\sqrt{f_{1}^{2}+f_{2}^{2}}$.
 The curves have been shown with three
values of free parameters $t_{\beta}$ where
$t_{\beta}=\tan\beta=f_{2}/f_{1}$.

From Fig.~(\ref{su3}-a), we can see that the contributions of the
$SU(3)$ simple group model to single top production are larger
than those of the LH model.  The $SU(3)$ simple group model has
negative contributions to single top production at the LHC and
the maximum deviation from SM is $7.5\%$ which is for the case of
$x_{\lambda}=4$ . For $f=1~\rm TeV$, $x_{\lambda}\geq3$, and
$1\leq t_{\beta}\leq5$, the absolute values of the relative
correction for s-channel, is in the ranges of $3.7\%\sim7.5\%$.
Fig.~(\ref{su3}-b) shows the relative correction
$R_W=\Delta\sigma(pp\rightarrow t\overline{b}X)$/$\sigma_{SM}$ as
a function of the mass new gauge boson $W_{X^-}$ in the context of
$SU(3)$ simple group model for $t_{\beta}$ and three values of
the mixing parameters $x_{\lambda}$.  The value of the relative
correction parameter $R_{X}$ is in the range of $ 0\%\leq
R_{X}\leq 0.15\%$. Thus, the effects of the new gauge boson
$W_{X}$ on the s-channel single top production might not be
detected at LHC.

\begin{figure}
\begin{center}
\centerline{\vspace{-1.2cm}}
\centerline{\includegraphics[scale=0.4]{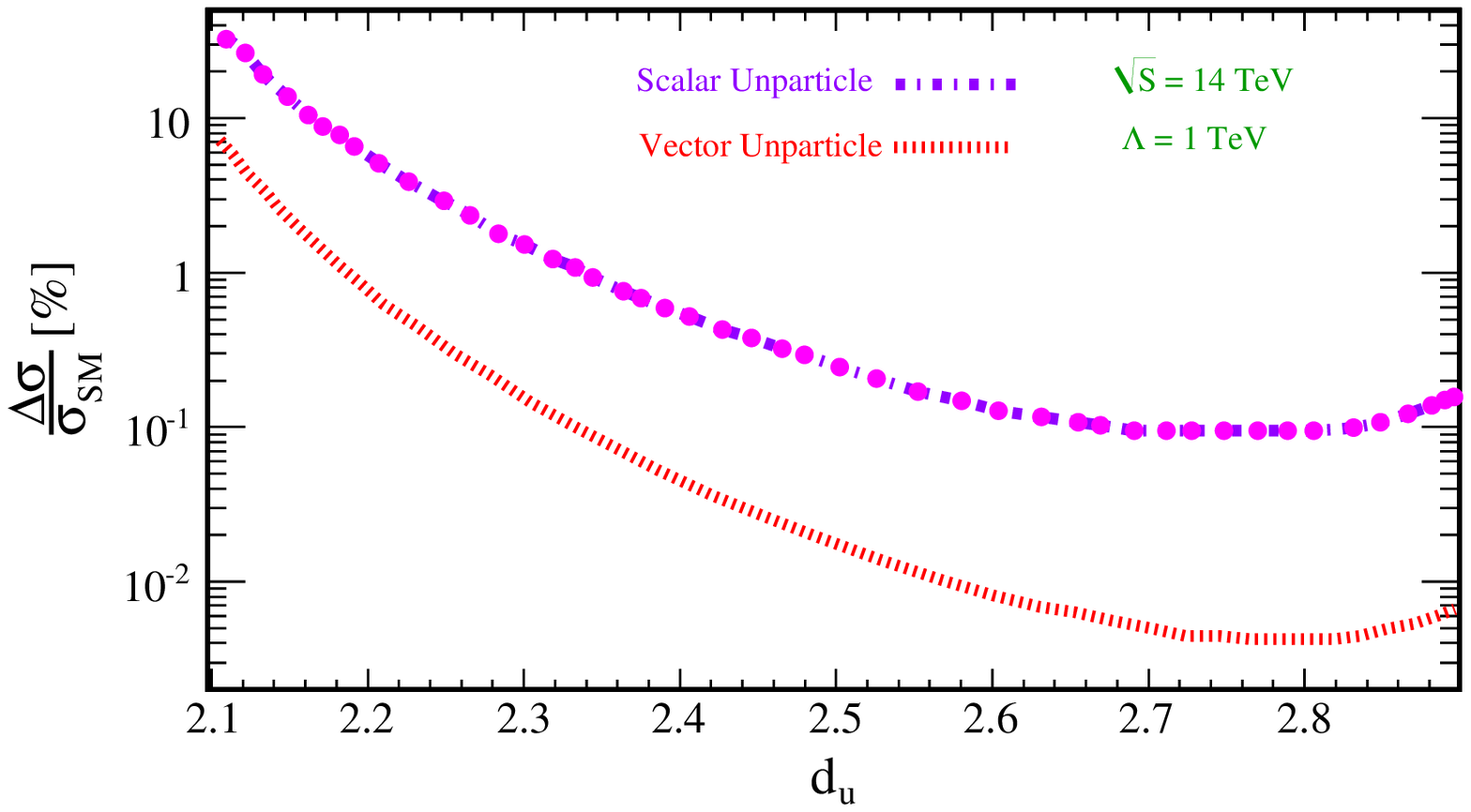}\hspace{0mm}\includegraphics[scale=0.4]{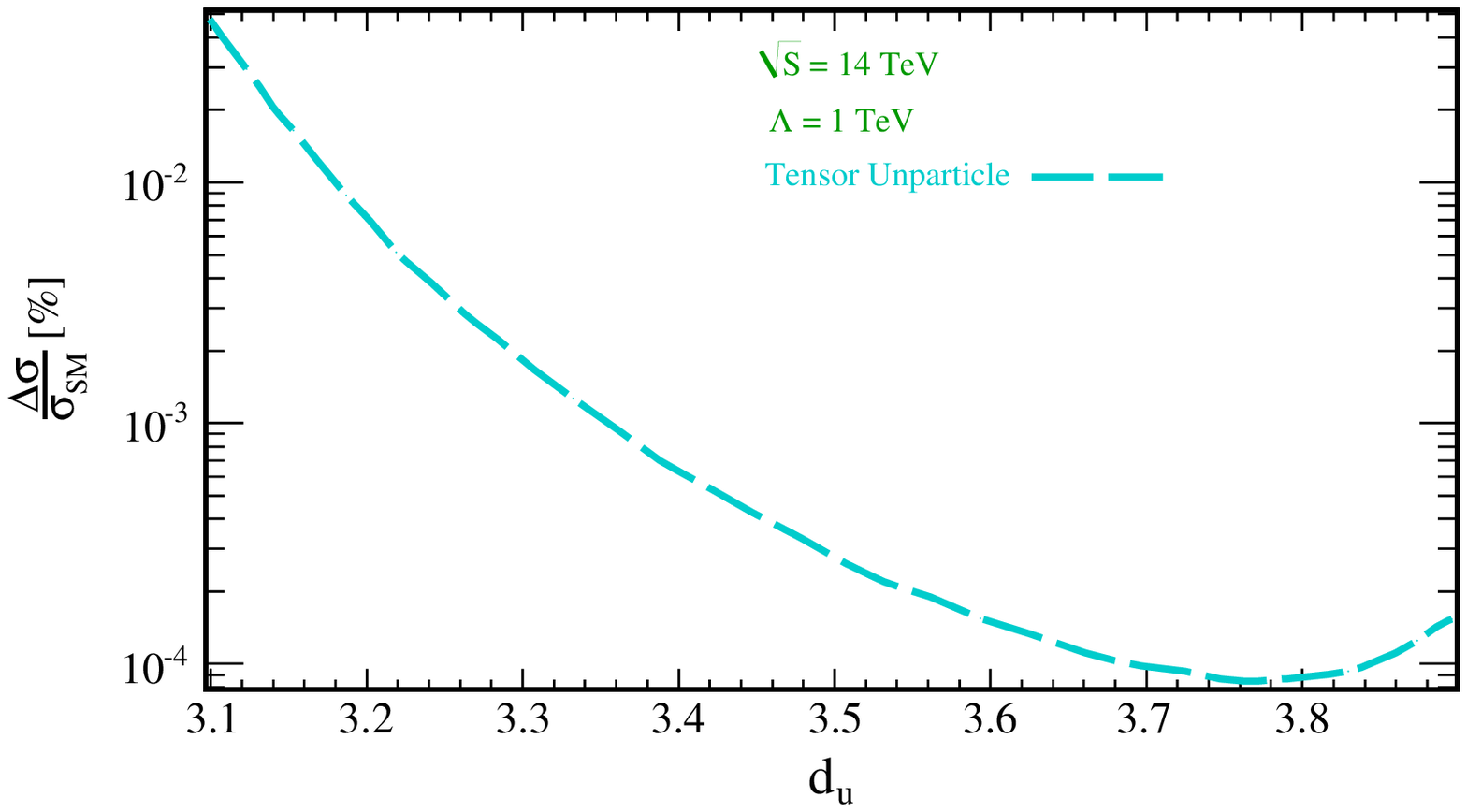}}
\centerline{\vspace{1.cm}\hspace{1cm}(a)\hspace{7cm}(b)}
\centerline{\vspace{-1.5cm}}
\end{center}
\caption{ a) $\Delta\sigma(pp\rightarrow
t\overline{b}X))$/$\sigma_{SM}$ versus scale dimension $d_u$ in
the context of unparticle physics for $\Lambda=1~\rm TeV$,
$\lambda_0=\lambda_1=1$ and $c_v=c_a$ at $14~\rm TeV$. The solid
violet curve corresponds to scalar unparticle and the dotted red
curve correspond to vector unparticle. b)
$\Delta\sigma(pp\rightarrow t+jet)$/$\sigma_{SM}$ versus the
scale dimension $d_u$ in the context of unparticle physics for
$\Lambda=1~\rm TeV$, $\lambda_2=1$ for tensor unparticle at
$14~\rm TeV$. Data for $\Delta\sigma(pp\rightarrow t+jet)$ have
been taken from Figs.~3 and 4 of
\cite{Unparticle}.}\label{Unparticle}
\end{figure}

 In the context of unparticle physics, SM fields can interact
with scalar, vector and tensor unparticles. The differential
cross-sections of single top production by considering these
three types of the unparticles have been presented in
\cite{Unparticle}. Contributions of unparticle physics to the
cross section $\sigma(pp\rightarrow t+jet)$ at the LHC are shown
in Fig.~\ref{Unparticle}. Data for $\Delta\sigma(pp\rightarrow t
+jet)$ have been taken from Figs.~3 and 4 of \cite{Unparticle}.

Figs.~\ref{Unparticle}-a and b, show $\Delta\sigma(pp\rightarrow t
+jet)$/$\sigma_{SM}$ versus the scale dimension $d_u$ in the
context of unparticle physics for $\Lambda=1~\rm TeV$,
$\lambda_0=\lambda_1=\lambda_2=1$ and $c_v=c_a$ at $14~\rm TeV$
where $\lambda_i$ are dimensionless effective couplings labeling
scalar, vector and tensor unparticle operators. $c_v$ and $c_a$
represent vector and axial vector coupling unparticle,
respectively. The solid violet curve corresponds to scalar
unparticle, the dotted red curve corresponds to vector unparticle
and the dashed cyan curve corresponds to tensor unparticles. With
the large integrated luminosity value at LHC, $L=10^5$ pb$^{-1}$,
we see that about 100 events are possible for both vector and
scalar mediated processes with $d_{u}=2.48$ and $d_{u}=2.34$,
respectively. As it is shown in Figs.~\ref{Unparticle},
$\Delta\sigma(pp\rightarrow t+jet)$/$\sigma_{SM}$ is always
positive and it is at most $20\%$, $6\%$ and $0.04\%$ for scalar,
vector and tensor unparticle (Note that SM leading order cross
section for s-channel cross section is $7~\rm pb$ at $14~\rm
TeV$). From these figures, we anticipate for the tensor and
vector unparticle  the cross sections are rather small for a wide
region of $d_{u}$ which we will not hope to detect them at LHC.
Since the change in cross section of s-channel due to scalar
unparticle is always positive, we can distinguish between extra
dimension with $KK$ excited $W$ gauge boson and unparticle
physics in large portion of parameters space.

\begin{figure}
\begin{center}
\centerline{\vspace{-1.2cm}}
\centerline{\includegraphics[scale=0.5]{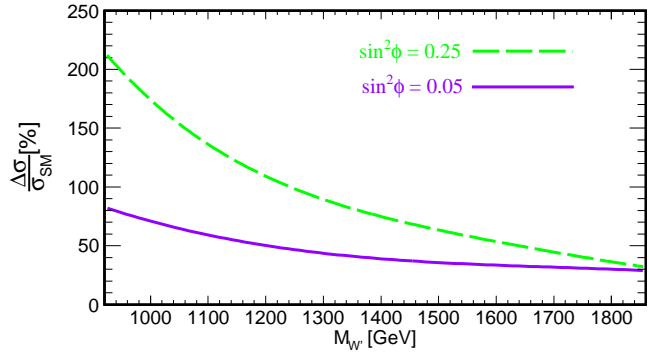}}
\centerline{\vspace{-1.5cm}}
\end{center}
\caption{ a) $\Delta\sigma(pp\rightarrow
t\overline{b}X)$/$\sigma_{SM}$ versus $M_{W'}$ in the context of
topflavor model for two values of the mixing parameters
$\sin^2\phi$. The solid violet curve corresponds to
$\sin^2\phi=0.05$ and the dashed green curve correspond to
$\sin^2\phi=0.25$. Data for $\sigma(pp\rightarrow
t\overline{b}X)$ have been taken from Figs.~7 of
\cite{Tait:2000sh}.}\label{topflavor}
\end{figure}

Another beyond SM theory which we are going to consider is
topflavor model \cite{topflavor, Tait:2000sh}. In context
topflavor model, the additional $W'$ boson can contribute to the
s-channel mode of the single top production through exchange of a
$W'$ boson. The relative corrections of the topflavor model to
the cross section $\sigma(pp\rightarrow t\overline{b}X)$ versus
$M_{W'}$ for two values of the mixing parameters $\sin^2\phi$ at
the LHC has been shown in Fig.~\ref{topflavor}. Data for
$\sigma(pp\rightarrow t\overline{b}X)$ have been taken from
Figs.~7 of \cite{Tait:2000sh}. In this figure, it is assumed that
$\Delta\sigma(pp\rightarrow t\overline{b}X)=\sigma_{\rm
topflavor}-\sigma_{SM}$. From Fig.~(\ref{topflavor}), we can see
that the contributions of the topflavor model to single top
production are very larger than other models. The topflavor model
has always positive contributions to single top production at the
LHC and the maximum deviation from SM is $210\%$ which is for the
case of $\sin^2\phi=0.25$. Thus, the effects of the new gauge
boson $W'$ on the s-channel single top production might be
detected at LHC and distinguishable is from our model.

The effect of the supersymmetric QCD corrections to the total
cross section for s-channel single top production at the LHC have
been presented in \cite{Supersymmetric}. It is shown that for
s-channel, the supersymmetric QCD corrections are at most about
$1\%$. Thus, we can ignore $\rm SUSY$ corrections in the future
high precision experimental analysis for s-channel single top
production at the LHC.

\section{Conclusions}
In extra dimension theories, if the gauge fields of the SM
propagate in the bulk of the extra dimension then they will have
KK excitations that can couple to the SM fermions. The masses of
KK excited gauge boson will be proportional to inverse radius of
compactification. In this paper, we have studied the impacts of
the first KK excitation of $W$ gauge boson on s-channel single
top production at LHC. We have shown that, if the inverse of
radius of compactification is in order of $\rm TeV$ scale or
less, phenomenological effects of these states at LHC will be
remarkable. The statistical uncertainty in the measurement of the
cross section for this process at the LHC will be about $5.4\%$
with an integrated luminosity of $30 fb^{-1}$\cite{1999fr}. It is
worthwhile to mention, the measurement of the cross section of
the s-channel single top at the LHC would not necessarily lead to
a deviation of $V_{tb}$ from one or evidence of new generation of
quarks. We have shown that for center of mass energy of $7~\rm
TeV$ if the mass of the first $W_{KK}$ is larger than $1650~\rm
GeV$, the total cross section of s-channel single top production
will be reduced and deviation from SM cross section can be up to
$10\%$ which indicates that effect of large extra dimension might
be detectable in this process. For center of mass energy $14~\rm
TeV$, this effect will be smaller and we need more integrated
luminosity than center of mass energy of $7~\rm TeV$.

To calculate s-channel single top production, we considered three
different parton distribution functions for quarks and showed
that deviation from SM cross section could not arise from these
effects.

We then compared our results with effects of other beyond SM
theories. It is shown\cite{Little Higgs} that for LH model and
topflavor model effects of new gauge boson on s-channel single top
production will be constructive. Thus degeneracy between these
models and our model will be distinguishable. We then discussed
the effects of QCD $\rm SUSY$ correction and $SU(3)$ simple group
model on single top production and mentioned these effects will
be very small and will not be detectable at LHC. For unparticle
physics model, effects of vector and tensor unparticle will be
small but scalar unparticle effect will be detectable at LHC.
Nevertheless this effect will be positive and so distinguishable
from effect of KK excitation of $W$ gauge boson.

\section*{Acknowledgement}
We would like to thank F.~Taghavi for helping in parton
distribution functions issues.

\end{document}